# Thermoelectrics: from Longitudinal to Transverse

Ken-ichi Uchida[1,2,]* and Joseph P. Heremans[3-5,]*

[1] National Institute for Materials Science, Tsukuba 305-0047, Japan

[2] Institute for Materials Research, Tohoku University, Sendai 980-8577, Japan

[3] Department of Materials Science and Engineering, The Ohio State University, Columbus, Ohio 43210, USA

[4] Department of Mechanical and Aerospace Engineering, The Ohio State University, Columbus, Ohio 43210, USA

[5] Department of Physics, The Ohio State University, Columbus, Ohio 43210, USA

*Correspondence: UCHIDA.Kenichi@nims.go.jp (K.U.), heremans.1@osu.edu (J.P.H.)

## Introduction

Thermoelectric power generation is a promising technology to realize a sustainable society because it directly converts thermal energy into electricity in a solid. Conventional thermoelectric generation is driven by the Seebeck effect, discovered by T. J. Seebeck in 1821, in which a charge current $\mathbf{J}_c$ is generated in the direction parallel to a temperature gradient $\nabla T$, the *longitudinal* geometry. The ratio between the generated longitudinal electromotive force and applied $\nabla T$ is defined as the Seebeck coefficient. As shown in Figure 1A, a thermoelectric module based on the Seebeck effect typically consists of many pairs (128 for a 12V unit) of *p*-type and *n*-type conductors alternately arranged and connected in series. Since the Seebeck coefficient of a *p*-type (*n*-type) conductor is positive (negative), the thermopower in each element adds to the total output in the Seebeck module. This configuration is needed to achieve a practical voltage because each element only supplies millivolts, but limits the use of thermoelectric generators. The efficiency of thermoelectric generators is characterized by the thermoelectric figure of merit, *zT*. Despite a three-fold improvement in *zT* resulting from materials research over the last two decades, little progress has been made in their practical use; this is ascribed to the technological challenges in contact technology.[1] The contacts on the hot side are subject to thermal degradation. All contacts (512 for a 12V unit) add electrical and thermal contact resistances that reduce the device efficiency to a fraction of that promised by the materials.

*Transverse* thermoelectric devices offer an alternative that avoids these problems.[2] Here, $\mathbf{J}_c$ is generated in the direction perpendicular to $\nabla T$. Now the voltage and power induced by transverse thermoelectric effects can be enhanced by increasing the length and area perpendicular to $\nabla T$, respectively, in a single material without forming a multielement structure (Figure 1B). The transverse thermoelectric effects can thus harvest thermal energy distributed over a large area. In the junctionless structure, no electrical contacts are needed on the hot side. The number of electrical contacts needed is reduced to two for extracting thermopower, bringing the efficiency of thermoelectric modules close to the theoretical value.[2]

In this paper, we first review various phenomena and mechanisms that produce the transverse thermoelectric effects and cite several studies that reveal their scientific and technological importance. We then discuss the potential of the transverse thermoelectric effects as "Future energy".

## Homogeneous Materials

*Nernst Effects*

The history of the transverse thermoelectric effects dates back to 1886, when A. V. Ettingshausen and W. Nernst discovered a phenomenon in which $\mathbf{J}_c$ is generated in the cross-product direction of $\nabla T$ and an external magnetic field $\mathbf{H}$ applied to a conductor.[3] This is called the Nernst or Nernst-Ettingshausen effect. Today, we often refer to this phenomenon as the ordinary Nernst effect (ONE) to distinguish it from the anomalous Nernst effect (ANE).

The ONE occurs in nonmagnetic conductors, due to the Lorentz force acting on free charge carriers (Figure 2A). Although the ONE in normal metals is small, Bi-based Dirac semimetals are known to exhibit very high transverse thermoelectric conversion performance due to the ONE (e.g., $zT > 0.3$ for BiSb alloys at ~ 1 T and 100-200 K).[4] However, an external magnetic field is required to implement this.

The ANE occurs in magnetic conductors, due to the spin-orbit interaction creating a Berry curvature in electronic bands and/or spin-dependent impurity scattering. It is related to the anomalous Hall effect (AHE) by the Mott relation: like the Seebeck coefficient is proportional to the energy derivative of the longitudinal electrical conductivity, the anomalous Nernst conductivity is proportional to the energy derivative of the anomalous Hall conductivity. Thus, mechanisms that give rise to the AHE give rise to the ANE. In ferromagnets, $\mathbf{J}_c$ is generated due to the ANE in the cross-product direction of $\nabla T$ and spontaneous magnetization $\mathbf{M}$ or the Berry curvature induced by the spin-orbit interaction (Figure 2B). Major recent developments for the ANE have been reported recently. Although the ANE-induced transverse thermopower, the anomalous Nernst coefficient, of pure Fe, Ni, and Co is in the order of 0.1 µV/K, the large coefficient of > 6 µV/K originating from topological electronic structures was observed in magnetic Heusler alloys, such as $Co_2MnGa$.[5] The large ANE has also been observed in binary Fe-based alloys,[6,7] kagome ferromagnet,[8] and rare-earth permanent magnets.[9] The record anomalous Nernst conductivity was reported in $YbMnBi_2$, a canted antiferromagnetic Weyl semimetal.[10] Because the ANE works even in thin films, it can be applied as a heat flux sensor with extremely small thermal resistance.[7] The ANE does not necessarily require an external magnetic field, since the remanent magnetization of magnetic materials can drive it. To realize energy harvesting and heat sensing applications of the ANE, further progress is needed since the anomalous Nernst coefficient is still much smaller than the Seebeck coefficient.

*Goniopolar Materials*

Transverse thermoelectric effects exist in homogenous anisotropic crystals even in the absence of a magnetic field and magnetization. In most single-crystalline conductors, the Seebeck coefficient has the anisotropy but the same sign irrespective of the crystallographic direction. Recently, the excellent transverse thermoelectric performance has been realized in "goniopolar" materials, where the Seebeck coefficients across different crystallographic directions have large magnitude and opposite signs. This is due to one of the following two mechanisms. First, in some metallic materials (e.g., $NaSnAs_2$), goniopolarity is induced by a topological feature of the Fermi surface in $k$-space: it must have a negative Gaussian curvature along one direction and a positive Gaussian curvature along another.[11] The Seebeck coefficient is negative along the direction of the positive curvature and positive along the other. The second mechanism is two-band conduction in semimetals and narrow-gap semiconductors at finite temperatures, where conduction and valence bands are occupied simultaneously, e.g., in $Re_4Si_7$.[2] When conduction electrons have a higher mobility along one direction while holes have a higher mobility along another, the thermopower along the former (latter) direction becomes negative (positive) due to the dominant electrons' (holes') contribution. The transverse thermoelectric generation is obtained by cutting a goniopolar crystal along the direction between the one that gives the largest *n*-type thermopower and the one that gives the largest *p*-type thermopower and by applying a temperature gradient along the direction that makes electrons and holes diffuse sideways (Figure 2C). Highly efficient transverse thermoelectric conversion with the transverse $zT \sim 0.7$ at 980 K has been reported in single-crystalline $Re_4Si_7$ without junction structures except for two electrodes located near the cold side of the sample used for extracting the thermopower.[2] The performance of the goniopolar material is outstanding compared to other principles, and will become the core of future transverse thermoelectric technologies (Figure 3). However, the choice of materials is presently limited, and further material exploration is desirable.

**Hybrid and Composite Materials**

*Spin Caloritronics*

Spin caloritronics is an interdisciplinary field based on the combination of spintronics and thermoelectrics, in which a spin current, a flow of spin angular momentum, is actively exploited as an energy carrier in addition to charge and heat currents. The rapid development of spin caloritronics was triggered by the discovery of the spin Seebeck effect (SSE) in 2008, which refers to the generation of a spin current from a heat current in a magnetic material.[12] A typical SSE device is a bilayer of a magnetic material and a normal metal (e.g., Pt) with strong spin-orbit interaction. The spin current generated by a temperature gradient in the magnetic material due to the SSE is converted into a charge current by the inverse spin Hall effect in the normal metal, enabling spin-current-driven transverse thermoelectric generation (Figure 2D). The SSE appears even in magnetic insulators, one of the unique features of the SSE.[13] In the early studies on the SSE, the ANE was a hindrance because the superposition of the ANE-induced thermopower makes it difficult to extract the pure spin-current-induced thermopower. To establish experimental techniques to separate the SSE from the ANE, both the phenomena were studied in parallel, resulting in renewed interest in the ANE. Now, the SSE and ANE can be distinguished from each other through the different thermoelectric conversion symmetries.[13] Although the thermoelectric output of the SSE is today still only of the order of the ANE, physics and materials science studies to improve the heat-spin and spin-charge current conversion efficiencies and to scale up from thin bilayer films to bulk composites are in progress.[14]

As an effort to further improve the transverse thermopower, a new transverse thermoelectric conversion mechanism appearing in magnetic metal/thermoelectric semiconductor hybrid materials has been proposed and demonstrated in 2021 (Figure 2E).[15] Since this phenomenon is driven by the artificial hybridization of the Seebeck effect in the thermoelectric semiconductor and the AHE in the magnetic metal, it is called the Seebeck-driven transverse thermoelectric generation (STTG). By optimizing the combination of the magnetic metal and thermoelectric semiconductor and their dimensions, the STTG can exhibit much larger transverse thermopower and $zT$ than those of the ANE, providing high flexibility for designing thermoelectric conversion properties. Based on a vast number of studies on the Seebeck effect and AHE, there is still plenty of scope for improving the performance of the STTG. However, the versatility of the STTG is less than that of the ANE; while the ANE can be respectively driven by in-plane and perpendicular temperature gradients in perpendicularly and in-plane magnetized configurations, the STTG can operate only in the perpendicularly magnetized configuration due to the symmetry of the AHE.

We note that the SSE and STTG, like the ANE, can operate even in the absence of a magnetic field if the magnetic material is uniformly magnetized. Therefore, for the zero-field operation of these phenomena, the optimal design of magnetic anisotropy is required; this is an important issue in spin caloritronics along with the improvement of the transverse thermopower.

### (p × n)-Type Transverse Thermoelectrics

In the artificial multilayer systems consisting of two different conductors obliquely and alternately stacked, anisotropic conduction of electrons and holes is introduced. When $p$-type carrier conduction appears in one direction and $n$-type in the orthogonal direction, the off-diagonal terms in the thermoelectric transport tensor become finite, resulting in the transverse thermoelectric conversion. Such artificial transverse thermoelectric materials are called the ($p \times n$)-type multilayers, which consist of millimeter-scale slab stacks or of nanometer-scale superlattices (Figure 2F).[16,17] The transverse thermopower and $zT$ in the ($p \times n$)-type multilayers can be optimized by selecting appropriate material components, engineering their band structures, and geometrically tuning the directions of charge and heat currents to the stacked structure. Significantly, the large transverse $zT$ of > 0.2 and thermoelectric cooling were observed in an alternately stacked BiSbTe-Bi system.[17] This approach also works without a magnetic field.

### Thermal Boundary Conditions

In longitudinal thermoelectrics, $zT$ does not depend on the geometry of the sample on which it is measured. This is not so for transverse thermoelectrics: the thermal boundary conditions along the transverse direction play an essential role. In a transverse thermoelectric device, a longitudinal heat flow can induce a transverse temperature gradient due to a thermal Hall effect; in other words, the thermal conductivity tensor, like the thermoelectric one, also has off-diagonal components. The transverse thermopower, electrical conductivity, and thermal conductivity are thus modulated depending on whether the transverse thermal boundary conditions are isothermal or adiabatic. In the isothermal condition, precautions have been taken to make the temperature gradient and heat flow parallel: the two sides of the sample where the voltage is measured are at the same temperature. In the adiabatic condition, the thermal Hall effect is allowed to develop, and no heat flux leaves the samples by the sides where the voltage is measured. These different thermal boundary conditions result in different definitions for the transverse $zT$. Unfortunately, the existing literature[18,19] uses contradictory definitions for the isothermal and adiabatic $zT$. This is discussed in a recent paper,[20] where it is experimentally demonstrated that the actual efficiency of transverse thermoelectric devices in the isothermal condition is higher than that in the adiabatic condition (note that, mathematically, the isothermal and adiabatic $zT$ can be converted into each other and the relation between the efficiency and $zT$ is essentially the same as that for the Seebeck effect).

### Towards Applications of Transverse Thermoelectrics

This paper reviews the transverse thermoelectric generation realized by various phenomena and principles. As summarized in Figures 2G and 3, each phenomenon is at a different research stage and has different characteristics. Goniopolar and ONE materials have given the excellent transverse $zT$ to date. However, the large ONE in Dirac semimetals works only at cryogenic temperatures and requires a strong magnetic field. Only one high-$zT$ goniopolar material, $Re_4Si_7$, is known. While the previous works established the proof of principle that transverse thermoelectric energy conversion is possible and while rhenium is an order of magnitude less expensive than precious transition metals, further material discovery is necessary. The spin-caloritronic phenomena are still at the fundamental research stage and the efficiency of the transverse thermoelectric generation using magnetic materials is currently very low, while spin caloritronics has produced new physics. To realize the application of spin caloritronics, it is necessary to develop and find materials with high heat-spin-charge current conversion properties. The recent fusion of spin caloritronics and topological materials science is a part of this effort. As exemplified by the discovery of the STTG and ($p \times n$)-type thermoelectric materials, developing hybrid/composite materials may be a breakthrough approach to enhance the transverse thermoelectric conversion performance. Furthermore, the hybrid thermoelectric generation based on the SSE and ANE is enabled in ferromagnetic metal/ferrimagnetic insulator junction systems[13] and ferromagnetic/nonmagnetic bulk nanocomposites.[14] As such hybrid/composite materials are yet to be optimized, there are lots of possibilities to improve the transverse thermoelectric conversion efficiency. Spin caloritronics, goniopolar materials, and ($p \times n$)-type thermoelectrics have never been combined, despite having similar functionality. The interdisciplinary fusion of these fields will bring innovation to transverse thermoelectrics, solving long-standing technological problems of thermoelectric generators.

**ACKNOWLEDGMENT**

The authors thank T. Hirai for the support on the figure preparation and T. C. Blackledge for the preliminary review of the manuscript. This work was partially supported by CREST "Creation of Innovative Core Technologies for Nano-enabled Thermal Management" (JPMJCR17I1) from JST, Japan (K.U.) and The U.S. Department of Energy grant DE-SC0020923 "Discovery of goniopolar metals with zero-field Hall and Nernst effect" (J.P.H.).

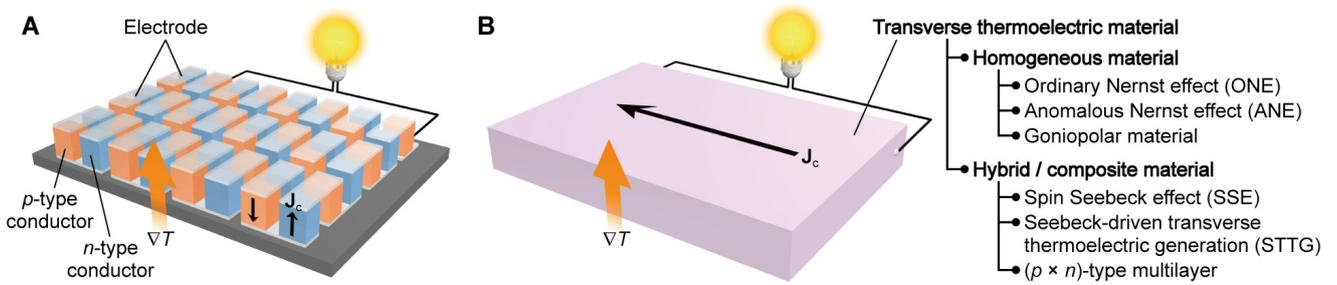

**Figure 1. Longitudinal and Transverse Thermoelectric Generation**

(A) Schematic of the thermoelectric module based on the Seebeck effect. (B) Schematic of the thermoelectric module based on the transverse thermoelectric effect. $\nabla T$ and $\mathbf{J}_c$ denote the temperature gradient and charge current driven by the thermoelectric effects, respectively.

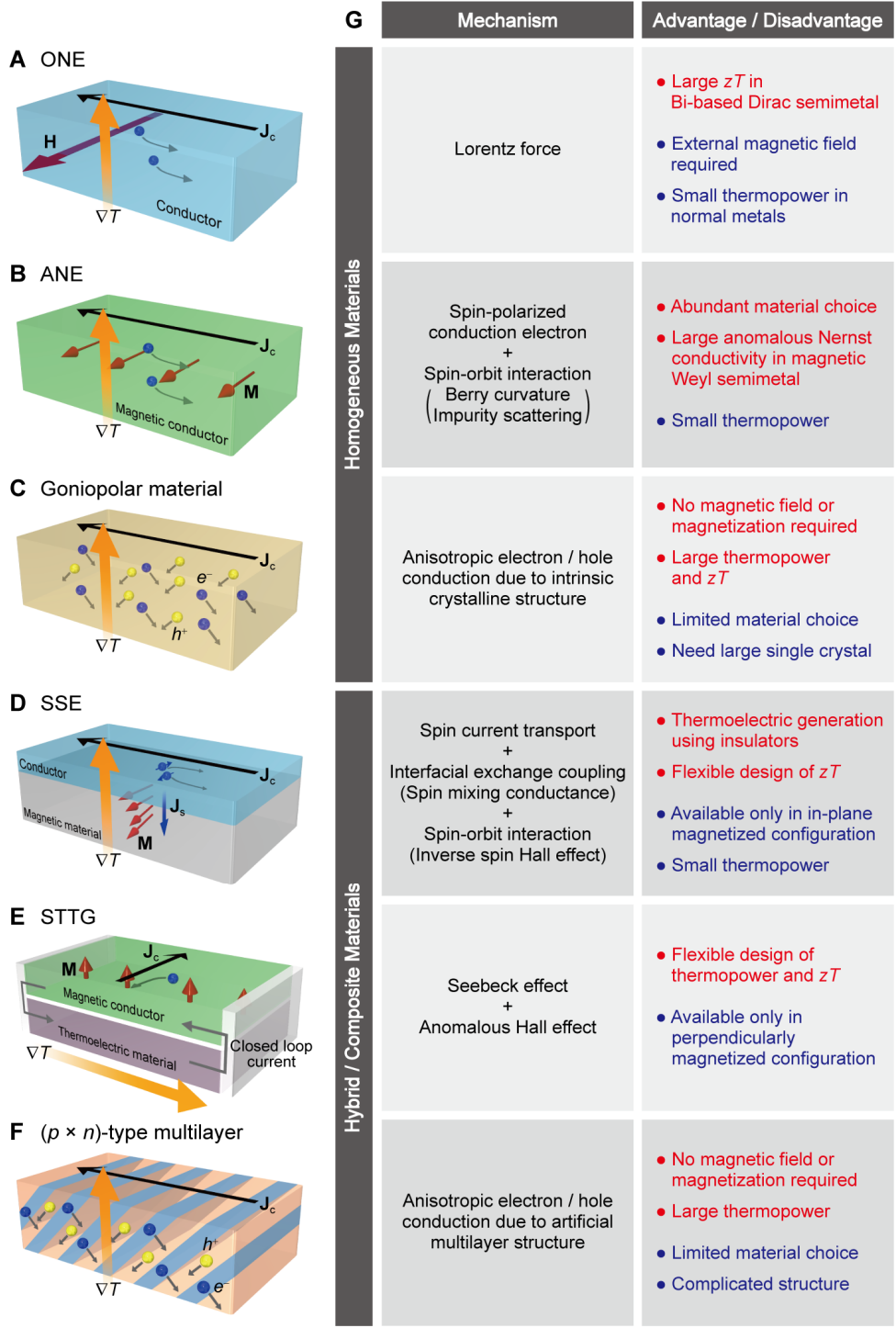

**Figure 2. Transverse Thermoelectric Effects**

Schematics of the ordinary Nernst effect (ONE) in a conductor (A), anomalous Nernst effect (ANE) in a magnetic conductor (B), goniopolar material (C), spin Seebeck effect (SSE) in a conductor/magnetic material junction system (D), Seebeck-driven transverse thermoelectric generation (STTG) in a magnetic/thermoelectric hybrid material (E), and ($p \times n$)-type multilayer (F). **H**, **M**, **J**$_s$, $e^-$, and $h^+$ denote the magnetic field, magnetization, spatial direction of a spin current, conduction electrons, and holes, respectively. The table in (G) summarizes the mechanisms and advantages/disadvantages of the transverse thermoelectric effects.

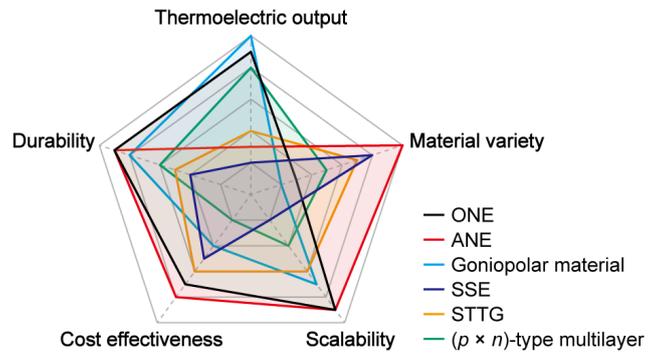

**Figure 3. Characteristic Comparison of Transverse Thermoelectric Effects**

The values in this radar chart are not quantitative, but rather guidelines determined based on the current research stage in order to compare the characteristics of each transverse thermoelectric effect.